\documentclass[conference]{IEEEtran}
\IEEEoverridecommandlockouts
\usepackage{cite}
\usepackage{amsmath,amssymb,amsfonts}
\usepackage{algorithmic}
\usepackage{graphicx}
\usepackage{textcomp}
\usepackage{xcolor}
\def\BibTeX{{\rm B\kern-.05em{\sc i\kern-.025em b}\kern-.08em
    T\kern-.1667em\lower.7ex\hbox{E}\kern-.125emX}}
\begin{document}

\title{A Study of Adoption and Effects of \\DevOps Practices}

\author{\IEEEauthorblockN{Tyron Offerman$^1$,
Robert Blinde$^1$, Christoph Johann Stettina$^1$, and Joost Visser$^1$}\\
\IEEEauthorblockA{$^1$LIACS, Leiden University, Niels Bohrweg 1, 2333 CA Leiden, The Netherlands\\}
}

\maketitle

\begin{abstract}

Many organizations adopt DevOps practices and tools in order to break down silos within the organization, improve software quality and delivery, and increase customer satisfaction. However, the impact of the individual practices on the performance of the organization is not well known. In this paper, we collect evidence on the effects of DevOps practices and tools on organizational performance. In an extensive literature search we identified 14 DevOps practices, consisting of 47 subpractices. Based on these practices, we conducted a global survey to study their effects in practice, and measure DevOps maturity. Across 123 respondents, working in 11 different industries, we found that 13 of the 14 DevOps practices are adopted, determined by 50\% of the participants indicating that practices are `always', `most of the time', and 'about half of the time' applied. There is a positive correlation between the adoption of all practices and independently measured maturity. In particular, practices concerning sandboxes for minimum deployment, test-driven development, and trunk based development show the lowest correlations in our data. Effects of software delivery and organizational performance are mainly perceived positive. Yet, DevOps is also considered by some to have a negative impact such as respondents mentioning the predictability of product delivery has decreased and work is less fun. Concluding, our detailed overview of DevOps practices allows more targeted application of DevOps practices to obtain its positive effects while minimizing any negative effects.
\end{abstract}

\begin{IEEEkeywords}
DevOps practices, DevOps tools, DevOps maturity, Software development, Organizational performance
\end{IEEEkeywords}

\section{Introduction}
More and more organizations are conducting digital transformation projects, leading them to increasingly rely on software~\cite{miller2006cots,ebert2018digital,gebhart2016challenges}. This puts pressure on their technology management capabilities, especially for organizations who decide to develop software themselves. Traditionally, the software development capability and software operation capability are separated from each other. This separation can be seen from a capability perspective, but also from an organizational unit or team perspective. Separating these two capabilities can hinder communication and collaboration.

DevOps was introduced to overcome this issue of misalignment and is concerned with problems organizations might encounter when releasing software in iterations. A key requirement in surmounting this misalignment requires organizations to change their culture to involve collaboration between different capabilities~\cite{ebert:2016}. While DevOps and the practice of continuous delivery can be used in tandem, DevOps also involves management and the required change in organizational culture~\cite{forsgren:2018}.
% Check for better 'bridge' between paragraphs

In this paper we study \emph{DevOps practices}, by which we mean a combination of practices from both capabilities and focus on DevOps' two main elements: cross-department collaboration and automation~\cite{leite:2019,ghantous:2017,riungu:2016}. Practices seen in DevOps are inspired by those found within Lean or Agile~\cite{forsgren:2018}. Many organizations adopt DevOps in order to break down the capability separation, improve software quality and delivery, and raise customer satisfaction~\cite{riungu:2016}. However, the impact of the individual practices on the performance of the organization is not well known. 

In this paper we report on our study, presenting an inventory of DevOps practices and their impact on DevOps maturity and organizational performance. To academia we provide further understanding of the impact of DevOps on organizations, explaining how DevOps practices develop through the different levels of maturity and how they impact organizational performance and software delivery performance. To practice we provide guidance on what DevOps practices to adopt to increase performance.

\section{Background and related work}
In this section we first highlight the history of DevOps and how it has been defined. Second, we provide an overview of DevOps practices found in literature.

\subsection{DevOps: history and definitions}
Today, with continuous changes in both market needs and technology, organizations cannot afford long lead times. The traditional waterfall method is therefore being replaced by new mindsets and methodologies: Agile and Scrum. While the focus was lying on increasing communication and collaboration between development teams and clients, the operations team -- in charge of deploying and maintaining the newest versions of the software -- was ``left out of the revolution''~\cite{hall:devops}. Still, within the waterfall method and Agile mindset organizations had separated their software development and operations teams~\cite{leite:2019}. DevOps was introduced to overcome the issue of the misalignment between development (capabilities) and operations (capabilities)~\cite{hall:devops}. We follow the definition of DevOps as proposed by Lwakatare et al.~\cite{lwakatare:2016}: ``DevOps is a mind-set substantiated with a set of practices to encourage cross-functional collaboration between teams - especially development and IT operations - within a software development organization, in order to operate resilient systems and accelerate delivery of change.''.

With the introduction of DevOps, multiple benefits after implementation of DevOps have been reported: automation of processes~\cite{wiedemann:2019,ghantous:2017}, shorter cycle times \cite{callanan:2016,riungu:2016,hamunen:2016,lwakatare:2019}, more frequent releases~\cite{catech:2013,senapathi:2018}, continuous experimentation and improvement~\cite{hamunen:2016,smeds:2015,riungu:2016,ghantous:2017}, increase in stability~\cite{hamunen:2016,forsgren:2019}, increase in quality~\cite{riungu:2016,callanan:2016,lwakatare:2019}, improved collaboration and communication~\cite{riungu:2016,shahin:2017}, and better and happier employees~\cite{senapathi:2018,smeds:2015}. 

As there is a lot of potential in the implementation of DevOps, there are also challenges such as: DevOps remaining a vague concept~\cite{smeds:2015,hamunen:2016,riungu:2016}, shifts in organizational culture~\cite{ebert:2016,riungu:2016,wiedemann:2019,ghantous:2017,smeds:2015}, lack of communication~\cite{riungu:2016,hamunen:2016}, management approval required~\cite{hamunen:2016,wiedemann:2019}, employees gaining new responsibilities~\cite{hamunen:2016,ghantous:2017,smeds:2015}, and DevOps being very context dependent~\cite{ebert:2016,riungu:2016,wiedemann:2019}.

An unclear definition of DevOps~\cite{ghantous:2017} or vagueness in the concept~\cite{smeds:2015,hamunen:2016,riungu:2016} can hinder its implementation. As there is no dominant DevOps definition in literature~\cite{smeds:2015}, we identified what the emphases of 17 DevOps' definitions are (seen in Tab.~\ref{tab:emphasis}). Here we do see common trends in the definitions. Most definitions focus, unsurprisingly, on development, operations, and delivery (teams) with a strong connection to the third emphasis of collaboration. The second emphasis is that DevOps is seen as a set of practices or capabilities. We also see the effects of DevOps returning in the definitions, such as reduction of time or release cycles or quality: correctness and reliability.

\begin{table}[t]
    \scriptsize
    \centering
    \caption{Emphases in DevOps definitions.}
    \label{tab:emphasis}
    \begin{tabular}{ll}
        \hline
        \textbf{Emphasis} & \textbf{Sources} \\
        \hline
        Development, operations, delivery (teams) &\cite{ebert:2016,jabbari:2016,ghantous:2017,lwakatare:2016,penners:2015,erich:2014,senapathi:2018,akshaya:2015,wettinger:2014,brunnert:2015,riungu:2016,erich:2017}\\
        Practices or capabilities & \cite{bass:2015,jabbari:2016,ghantous:2017,smeds:2015,lwakatare:2016,senapathi:2018,akshaya:2015} \\
        & \cite{balalaie:2016,brunnert:2015} \\
        Collaboration &\cite{yasar:2016,jabbari:2016,leite:2019,lwakatare:2016,penners:2015,senapathi:2018,akshaya:2015}\\
        & \cite{erich:2017} \\
        Quality: correctness and reliability & \cite{leite:2019,ghantous:2017,lwakatare:2016,penners:2015,balalaie:2016,brunnert:2015,riungu:2016}\\
        Reduction of time or release cycles&\cite{bass:2015,yasar:2016,lwakatare:2016,balalaie:2016,riungu:2016} \\
        Framework, principles, approach&\cite{erich:2014,senapathi:2018,brunnert:2015,riungu:2016}\\
        Mindset&\cite{lwakatare:2016,penners:2015}\\
        Communication&\cite{yasar:2016,riungu:2016}\\
        Continuous everything&\cite{jabbari:2016,leite:2019}\\
        Automation&\cite{jabbari:2016,leite:2019}\\
        Culture&\cite{smeds:2015}\\
        Technology enablers&\cite{smeds:2015}\\
        Lean&\cite{ebert:2016}\\
        Silos&\cite{ebert:2016}\\
        Paradigm&\cite{wettinger:2014}\\
        \hline
    \end{tabular}
\end{table}

\subsection{DevOps practices}
Definitions give some insights into what is meant by DevOps, as seen in Tab.~\ref{tab:emphasis}. Although some scholars have attempted to clarify DevOps concepts and practices~\cite{smeds:2015}, it still remains unclear how DevOps can be adopted effectively. In order to make it more explicit, it is helpful to dive deeper into the practices of Dev\-Ops. In total 47 DevOps practices were found in literature. The most common DevOps practices can be seen in Tab~\ref{tab:PracLiterature}. Here we see practices that are concerned with the dev(elopment) and op(erations) part of DevOps, as well as practices which provide an integrated view.

\begin{table*}[t]
    \scriptsize
    \centering
    \caption{Most common DevOps practices.}
    \label{tab:PracLiterature}
    \begin{tabular}{lll}
        \hline
        \textbf{\#} & \textbf{Practice} & \textbf{References} \\
        \hline
        1 & Automated and continuous deployment throughout entire pipeline & \cite{jabbari:2016,aljundi:2018,lwakatare:2019,humble:2018,ghantous:2017,senapathi:2018,stahl:2017} \\
        2 & Make small and continuous releases & \cite{stahl:2017,hamilton:2007,patel:2009} \\
        3 & Developers get feedback based on releases & \cite{hamilton:2007,patel:2009} \\
        4 & Create development sandboxes for minimum code deployment & \cite{ghantous:2017} \\
        5 & Everything is stored as code and under version control & \cite{ghantous:2017,hamilton:2007,humble:2018,jabbari:2016,lwakatare:2019} \\
        6 & Integrated configuration management & \cite{jabbari:2016} \\
        7 & Automated and continuous testing in development and staging environments & \cite{jabbari:2016,patel:2009,ghantous:2017,humble:2018,hamilton:2007} \\
        8 & Reduce the time it takes to test, validate and QA code & \cite{aljundi:2018} \\
        9 & Code reviews are change based & \cite{lwakatare:2019} \\
        10 & Automated and continuous monitoring of applications and resources & \cite{jabbari:2016,lwakatare:2019,hamilton:2007,balalaie:2016} \\
        11 & Automated dashboards that include health checks and performance & \cite{jabbari:2016,hamilton:2007} \\
        12 & Support configurable logging that can optionally be turned on/off as needed & \cite{hamilton:2007} \\
        13 & Use trunk-based development over long-lived feature branches & \cite{lwakatare:2019,humble:2018} \\
        14 & Use Test driven Development where all code has unit tests & \cite{patel:2009} \\
        \hline
    \end{tabular}
\end{table*}

\subsection{Lens of capabilities}
In order to breakdown DevOps into its practices, we apply the scientific lens of capabilities to it. Capabilities have been seen before as a source of competitive advantage~\cite{teece1997dynamic,zafer2009development}. In 2009, Salvato\cite{salvato:2009} analyzed the how organization-wide capabilities are linked to day-to-day routines/practices. He found that these day-to-day routines provide the basis for sustaining the competitive advantage of a company. Salvato~\cite{salvato2009capabilities} recommends, based on his findings, to shift focus from capabilities towards more fine-grained micro-activities as they provide more detailed evidence on the source of competitive advantage. Salvato and Rerup\cite{salvato2011beyond} also argues more multilevel research on (organizational) routines and capabilities is necessary. Teece\cite{teece2018business} also emphasizes the importance of further investigating the different levels of capabilities and their linkage to an organization's business model. By taking a multilevel approach on capabilities, we better understand the breakdown of a capability. We refer to this as the unboxing of a capability.

\section{Research question}
% As seen, high levels of automation, collaboration, and correct tools for the job lead to shorter lead times.
As seen, implementing DevOps practices has both benefits and challenges. However, it remains unclear whether these practices influence the performance of organizations. Therefore, we pose the following research question: \textit{What are the effects of DevOps practices on an organization’s performance?} To answer this question regarding the effects of DevOps practices, we also dive deeper in into which DevOps practices are adopted to which degree by various organizations. Answering this question contributes to both academia and practitioners: we provide 1) further understanding of the impact of DevOps on organizations and 2) guidance on what DevOps practices to adopt.
%outline/contribution

\section{Methodology}
To understand the DevOps practices and their effects on organizations, data from a wide range of DevOps professionals was necessary.
%%%
Thus, we set up a survey.

\subsection{Survey design}
The survey consists of following 6 categories: (1) context, (2) transformation, (3) maturity models, (4) practices and tools, (5) impact of DevOps, and (6) organizational performance.

(1) Context - The first category of the survey gathers descriptive information to verify the spread of the sample. This category includes questions regarding personal background (role and years of experience with DevOps) and organizational context (industry and size of the organization). These questions are used to group the participants.

(2) Transformation - The second category of the survey is focused on the DevOps transformation. Here questions are asked whether the participant's organization has gone through, is in the middle of, or is planning a DevOps transformation. If this is the case, we pose questions about the transformation itself and what frameworks have supported the transformation.

(3) Maturity Models - Third, the perceived Agile and DevOps maturity of the organizations was asked. For Agile maturity the questions are based on Laanti's Agile Maturity Model~\cite{laanti:2017}.  Here participants rate their organization's maturity on three organizational levels: portfolio, program, and team. Each organizational level is measured as `Beginner', `Fluent', or `World-class'. To measure DevOps maturity questions are based on maturity model by Eficode~\cite{eficode:2020}. This DevOps maturity model consists of six organizational areas: Organization \& Culture, Environments \& Release, Builds \& Continuous Integration, Quality Assurance, Visibility \& Reporting, and Technologies \& Architecture. Each organizational area is measured according to a scale ranging from level 1 to level 4. The questions for both Agile and DevOps maturity have been accompanied by the complete maturity model in order to give participants context on how to measure.

(4) Practices and tools - In the literature review, we identified common Dev\-Ops practices (Tab.~\ref{tab:PracLiterature}) and technologies used for each of these practices. In the fourth category of the survey we measure these practices, and their sub-practices. For each sub-practice, the participants are asked how often they adhere to each practice based on a 7-point Likert scale. If at least one of the sub-practices of a practice is adhered to, the participants are asked to identify which technologies are used to support the practice.

(5) Impact of DevOps - The impact of DevOps on software delivery, on the participants and their team members, and on customer satisfaction is the focus of the fifth category of the survey. To measure the impact on software delivery we applied the construct for Software Delivery Performance as defined by Forsgren et al.~\cite{forsgren:2018}. The construct consists of four questions regarding lead time, deployment frequency, mean time to restore, and change fail percentage. The measurement of the impact of DevOps on the participants and their team members was conducted by asking about the following metrics, as adopted from Laanti~\cite{laanti:2017}: effectiveness of development, quality of the product, customer satisfaction, collaboration, work being more fun, work being less hectic, work being more organized, and earlier detection of bugs/errors. The customer satisfaction metric is extended by questions such as the usefulness of the product, usability of the product, and predictability of delivery. The metrics are measured by the percentage of change the impact has, ranging from -100 (negative impact) to 100 (positive impact).

(6) Organizational Performance - The last category of the survey is centered around organizational performance. Participants are questioned to rate their organization's relative performance on the following dimensions: overall performance, overall profitability, customer satisfaction, quality of products and services, operating efficiency, and achieving organizational goals. The questions to measure these dimensions are adopted from Widener~\cite{widener:2007}. The answers are on a 7-point Likert scale, ranging from `Performed well above goals' to `Performed well below goals'.

\subsection{Data collection / distribution}
Our survey was distributed using a snowball strategy. First, we approached our own network via LinkedIn and personal contacts. Second, we approached (online) DevOps communities and organizations to distribute the survey. Organizations were approached based on whether or not they develop software. Selection criteria to distribute the survey via the organizations were if they had implement DevOps practices or tools.

\section{Results}
In the period from June 30 2021 to November 5 2021, 337 people started the survey and 123 completed it, resulting in a response rate of 36.5\%.

\subsection{Participants and their organizations}
Roles and experience: The most common roles of the participants were the roles of software developer/engineer (30.1\%) and IT operations/infrastructure engineer (22\%). Other roles, with a higher percentage than 4\%, were automation engineer/expert (8.9\%), product manager (5.7\%), and project manager (4.1\%). Most participants (38.2\%) have 3 to 5 years of experience in/with DevOps, followed by 1 to 2 years (22.0\%), and 6 to 10 years (17.9\%).

Industry and organization: The participants are working primarily in the technology sector (32.5\%), followed by financial services (15.5\%), and retail, consumer, \& e-commerce (8.1\%). Throughout the different industries, we see participants commonly working in organizations with 20 to 99 employees (24\%), 100 to 499 employees (20.7\%), or 10,000+ employees (20.7\%).

Transformations: Most of the organizations have completed a DevOps or Agile transformation (46.3\%) or are currently undergoing a transformation (39.2\%). Some are about to start such a transformation (4.9\%), while 9.8\% of the participants have indicated that they are not planning to go through a transformation or have not completed any. Participants who completed or are undergoing a transformation were asked which frameworks they use to support the transformation. We found that 60\% of the participants did not use a DevOps framework during the transformation. The most applied Dev\-Ops frameworks were CALMS (20.9\%) and SAFe's CALMR (11.8\%). The most common Agile frameworks were Scaled Agile Framework (26.1\%), Enterprise Scrum (22.5\%), Scrum of Scrums (16.2\%), and Agile Portfolio Management (14.4\%). There was also a group of participants (11.7\%) who indicated to use an internally created methodology, where 9\% of the participants indicated to not use an Agile framework during the transformation.

\subsection{Adoption of DevOps practices}
\begin{figure}
  \centering
  \includegraphics[scale=0.4]{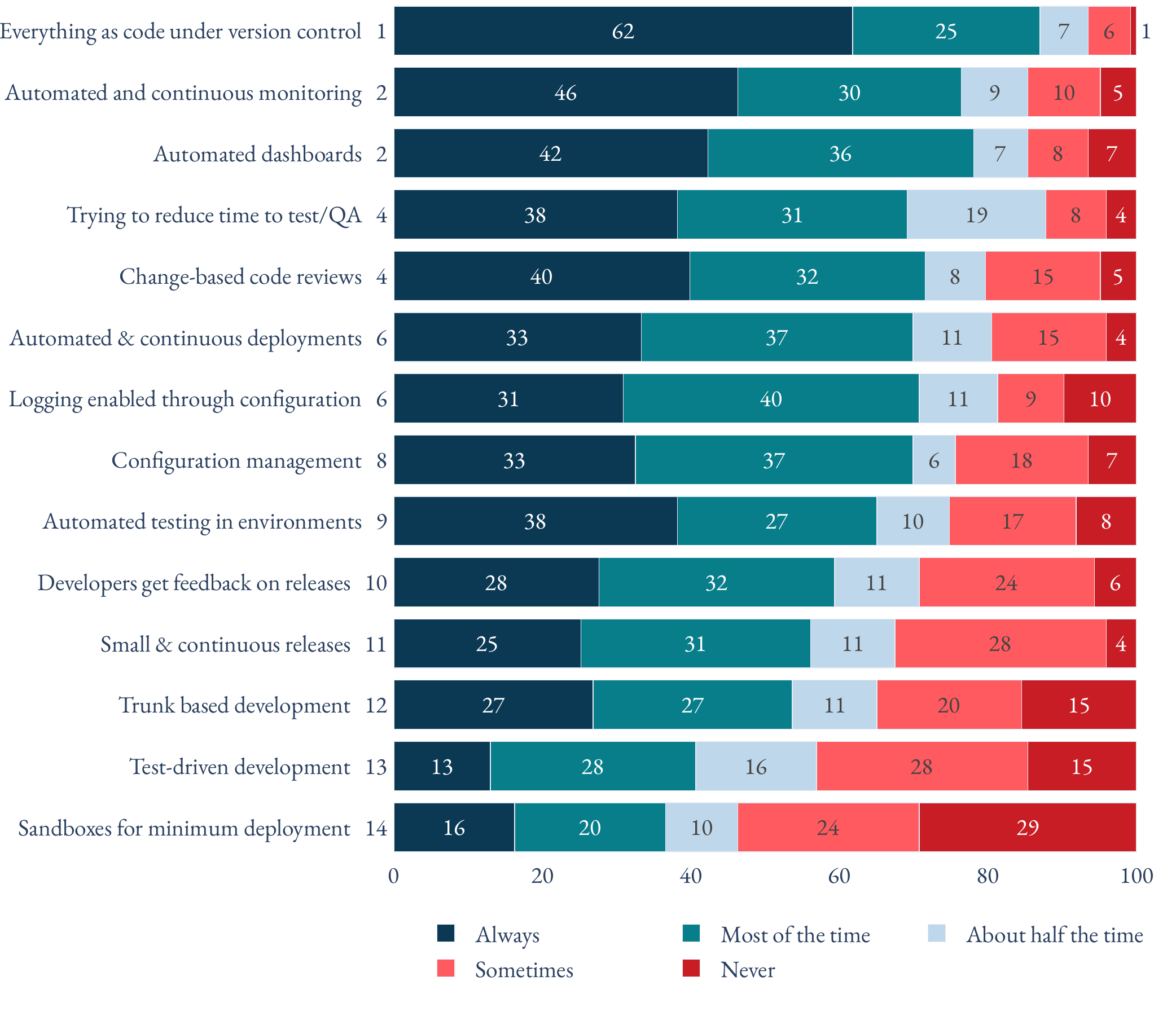}
  \caption{Usage of DevOps practices.}
  \label{fig:practice_usage}
\end{figure}

To measure the adoption of DevOps practices, participants were asked to indicate how often they adhere to DevOps practices. Fig.~\ref{fig:practice_usage} shows the usage of DevOps practices. For each of the following DevOps practices, participants indicated that they always apply the practice: `Everything as code under version control' (62\%), `Automated testing in environments' (38\%), `Trying to reduce time to test/QA' (38\%), `Change-based code reviews'(40\%), `Automated and continuous monitoring' (46\%), `Automated dashboards' (42\%), and `Trunk-based development' (27\%). For the latter practice their is an equally large group who indicate that they apply the practice most of the time.

This is followed by the DevOps practices, which have been indicated to be applied most of the time by the largest group: `Automated \& continuous deployments' (37\%), `Small \& continuous releases' (31\%), `Developers get feedback on releases' (32\%), `Configuration management'(37\%), `Logging enabled through configuration' (40\%), `Trunk-based development' (27\%), `Test-driven development' (28\%). For the latter practice their is an equal large group who indicate that they apply the practice sometimes. Most of the participants (29\%) indicated to never apply the practice`sandboxes for minimum deployment'.

Additionally, we calculated the adoption rate based on the ranking algorithm, as described by Serban et al.~\cite{serban2020adoption}, with the following steps:
\begin{enumerate}
\item For each practice we calculate the percentage of respondents who responded with at least high adoption (answering `always'), the percentage with at least medium adoption (answering `always' or `most of the time'), and the percentage with at least low adoption (answering `always', `most of the time', and `about half of the time').
\item The percentages are transformed into ranks.
\item We take the average of the three ranks and then rank the averages in reserve order. The lower the rank the better the adoption of the DevOps practice.
\end{enumerate}
The results can also be seen in Fig.~\ref{fig:practice_usage}. Here we see `everything as code under version control' as the most adopted practice, followed by `automated and continuous monitoring' and `automated dashboards'. The least adopted practice is `sandboxes for minimum deployment'.

\subsection{Impact and maturity of DevOps}
%\begin{figure}[!t]
 % \centering
  %\includegraphics[scale=0.5]{images/performance.png}
  %\caption{Organizational performance across six aspects (as from~\cite{widener:2007}).}
  %\label{fig:org_perform}
%\end{figure}

Participants were asked to indicate how well their organization performed across six dimensions as defined by Widener~\cite{widener:2007}. These six dimensions are as follows: performance, profitability, customer satisfaction, quality, efficiency, and achieving goals. The answers are on a 7-point Likert scale, ranging from `Performed well above goals' to `Performed well below goals'. For each of the dimensions 50\% to 60\% of the participants indicated that their organization `performed above goals' or `met goals', with 80\% of the participants indicating the organizations met their goals or better. In less than 15\% of the cases the performance of the organization was rated to perform below goals or lower.

\begin{figure}[t]
  \centering
  \includegraphics[scale=0.38]{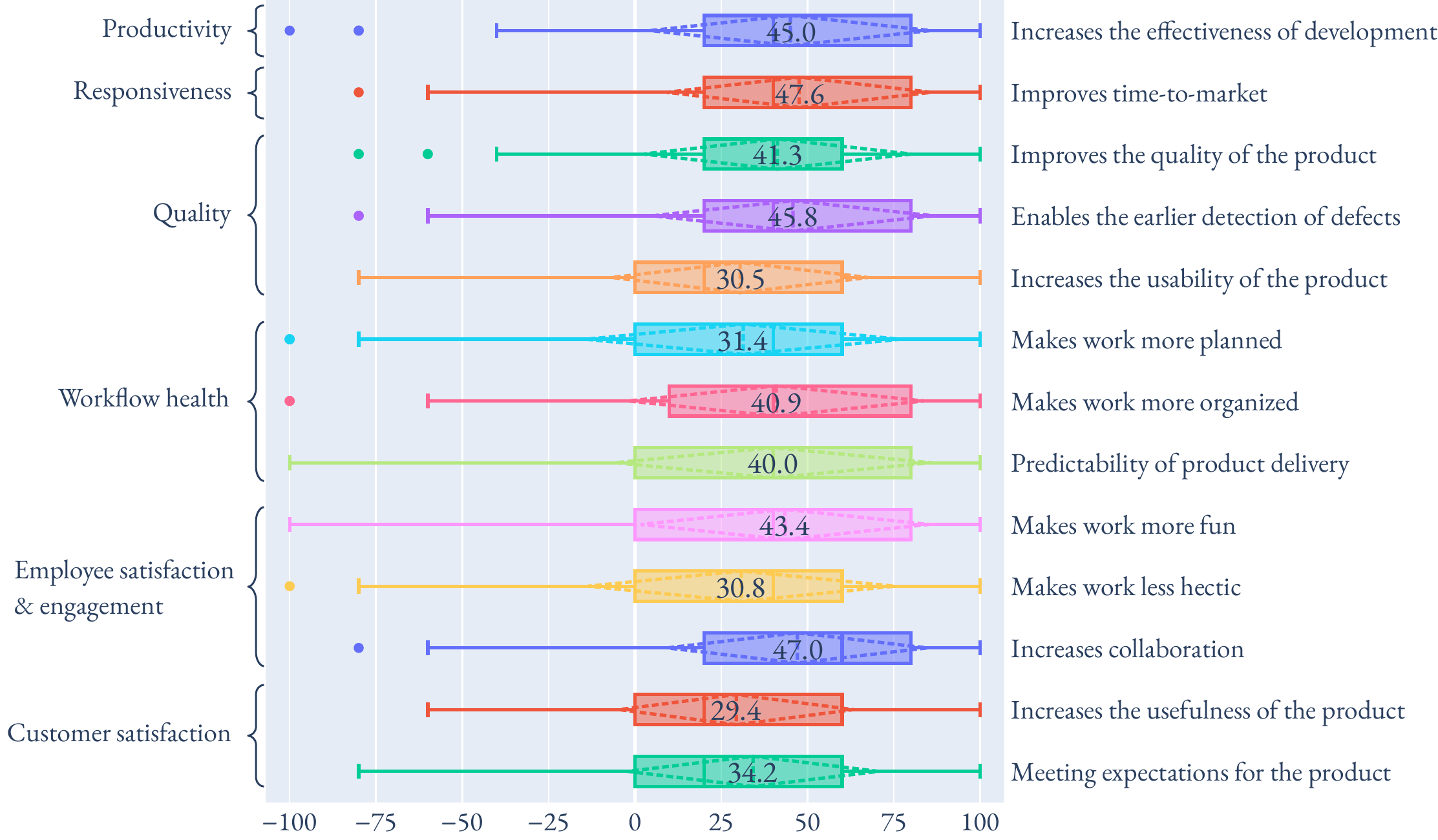}
  \caption{DevOps impact on thirteen dimensions (adopted from~\cite{stettina:2021}).}
  \label{fig:impact}
\end{figure}

Fig.~\ref{fig:impact} displays the perceived organizational impact of a DevOps implementation, categorized into six dimensions. On average, all thirteen aspects show a positive impact. However, for all dimensions, there are participants who have indicated negative impacts. The aspects with the highest perceived impact are: `improves time to market' (47.6\%), `increases collaboration' (47.0\%), `enables the earlier detection of defects' (45.8\%), and `increases the effectiveness of development' (45.0\%). `Increases the usefulness of the product' (29.4\%), `increases the usability of the product' (30.5\%), `makes work less hectic' (30.8\%), and `makes work more planned' (31.4\%) were the aspects with the lowest perceived impact. 25\% of the respondents experienced a negative impact (-100 to 0) for the aspects `makes work more fun' and `increases predictability of product delivery'.

\begin{figure}[t]
  \centering
  \includegraphics[scale=0.45]{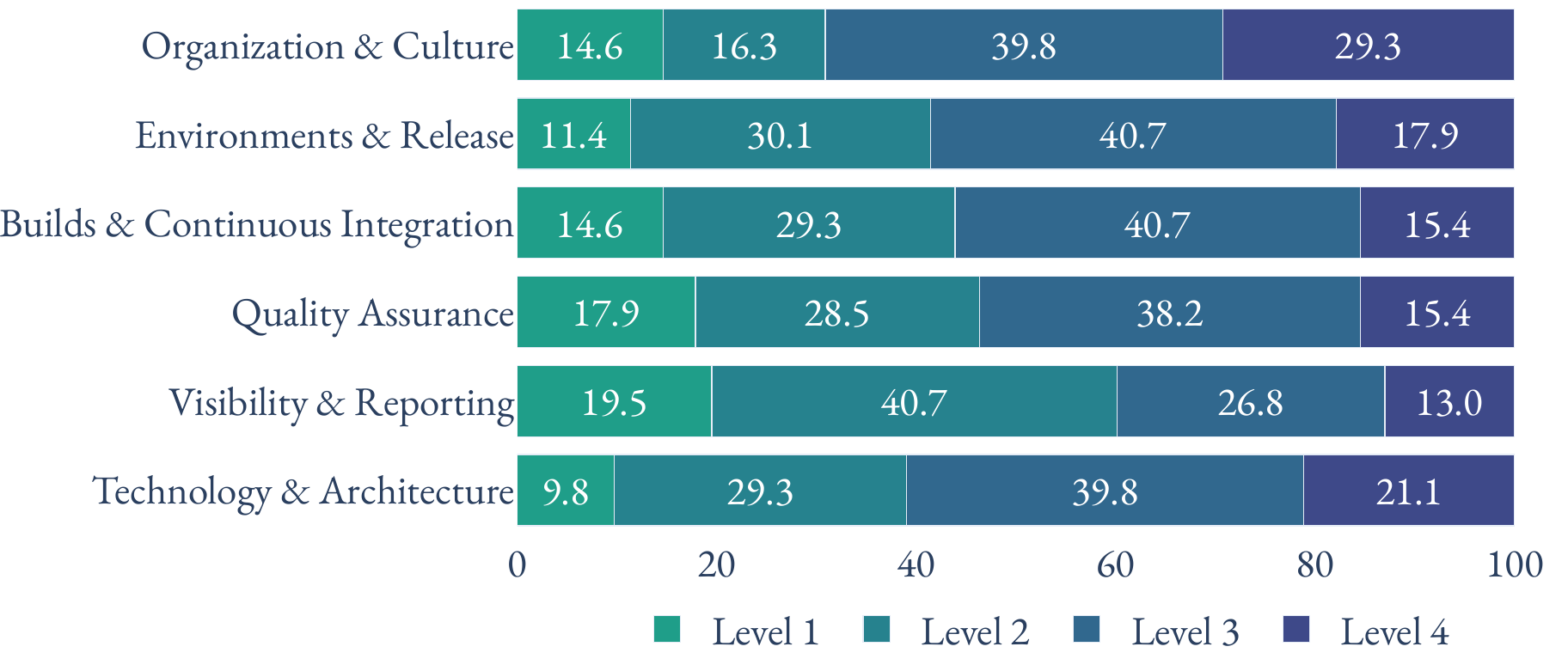}
  \caption{DevOps maturity across six aspects.}
  \label{fig:devops_maturity}
\end{figure}

Fig.~\ref{fig:devops_maturity} shows the DevOps maturity as indicated by the participants. The figure shows the six DevOps maturity aspects as defined by Eficode~\cite{eficode:2020} across level 1 to 4. For the aspect `organization \& culture' the largest group of participants assessed their organization as level 3 (40\%). Level 4 was the second largest group with 29\% for this aspect, followed by level 2 (16\%) and level 1 (15\%). Aspect `environments \& release' is assessed at level 3 by most of the participants (41\%). 30\% of the participants assessed their organization at level 2, followed by level 4 (18\%) and level 1 (11\%). The third aspect of the DevOps maturity model is `builds \& continuous integration'. The largest group of participants can be seen at level 3 (41\%). Level 2 is indicated in 29\% of the cases, while both level 1 and level 4 have been indicated in 15\% of the cases. `Quality Assurance' is assessed at level 3 by 38\% and at level 2 by 28\% of the participants. Level 1 is mentioned by 18\% of the participants, followed by level 4 (15\%).  At `visibility \& reporting' we see level 2 as the largest level (41\%), followed by level 3 at 27\%, level 1 at 20\%, and level 4 at 10\%. The last aspect of the DevOps maturity model is `technology \& architecture' where level 3 (40\%) is the largest and level 2 (29\%) the second largest. Level 4 was assessed by 21\% of the participants and level 1 by 10\%.

\subsection{Correlations}
Measuring the effects of DevOps practices on organizational performance requires analysis of their correlations.
%Figures~\ref{fig:cor_prac_perf} and~\ref{fig:cor_prac_sdp} show the correlation between Dev\-Ops practices and performance or software delivery performance, respectively.

\subsubsection{Organizational performance}
\begin{figure*}
  \centering
  \includegraphics[scale=0.64]{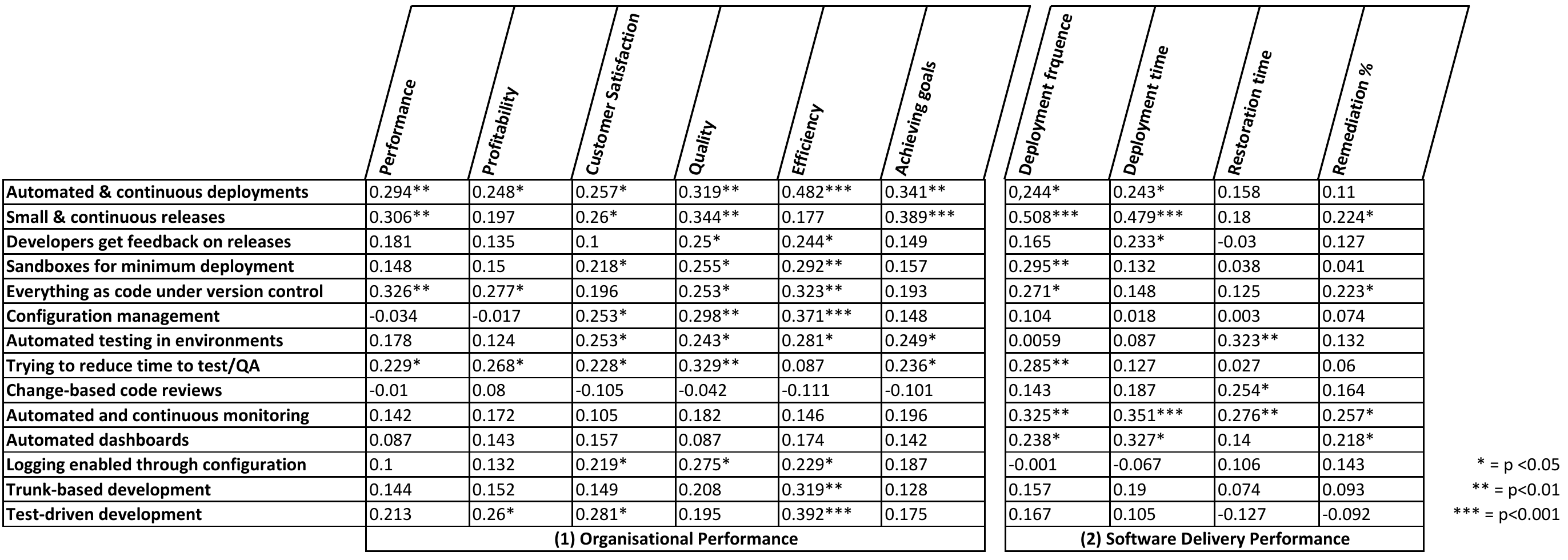}
  \caption{Correlations between DevOps practices and (1) organizational performance (left) and (2) software delivery performance (right).}
  \label{fig:cor_prac_perf}
\end{figure*}

When correlating DevOps practices and organizational performance (Fig.~\ref{fig:cor_prac_perf} left), we found the following correlations to be the strongest: `automated continuous deployments' (corr.: 0.248 - 0.482), `small \& continuous releases' (corr.: 0.177 - 0.389), and `test-driven development' (corr.: 0.175 - 0.392). The weakest correlations were found at the practices: `change-based code reviews' and 'automated dashboards'. For organizational performance we found 7 negative correlations. The DevOps practice `change-based code reviews' had the most (5) negative correlations, with only `profitability' having a small positive correlation. Other negative correlations were found for the practice `configuration management'.

\subsubsection{Software delivery performance}
The same figure (Fig.~\ref{fig:cor_prac_perf} right) also includes correlations between DevOps practices and software delivery performance.
Here, we found the following strongest correlations: `small \& continuous releases' (corr.: 0.18 - 0.508), `automated and continuous monitoring' (corr.: 0.257 - 0.351), and `automated dashboards' (corr.: 0.14 - 0.327). The weakest correlations were found for the following practices: `configuration management', `logging enabled through configuration', and `test-driven development'. Also in the correlations between DevOps practices and software delivery performance, we found negative correlations. Five negative correlations were for the practices `developers get feedback on releases', `logging enabled through configuration', and `test-driven development'.

\subsubsection{Maturity}
\begin{figure}
  \centering
  \includegraphics[scale=0.5]{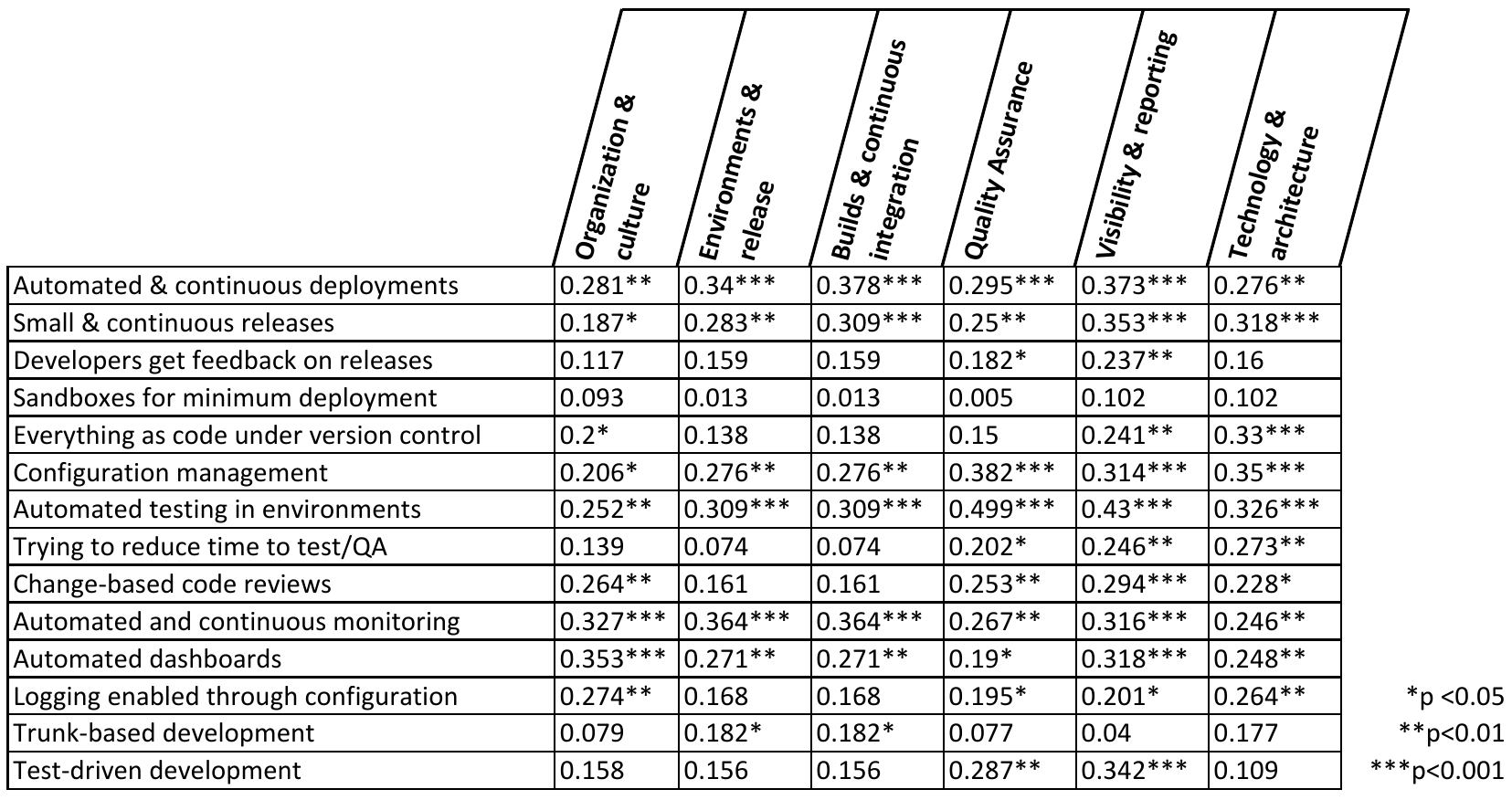}
  \caption{Correlation between DevOps maturity and practices.}
  \label{fig:cor_dm_prac}
\end{figure}

In Fig.~\ref{fig:cor_dm_prac} we provide an overview of the correlations of the aspects of DevOps maturity per DevOps practice usage. The strongest correlations are as follows: `automated testing in environments' with correlations between 0.252 and 0.499; `automated continuous deployments' with correlations between 0.276 and 0.378; `automated and continuous monitoring' with correlations between 0.246 and 0.364; `configuration management' with correlations between 0.206 and 0.382. `Sandboxes for minimum deployment' with correlations between 0.005 and 0.102; `trunk-based development' with correlations between 0.04 and 0.182; `trying to reduce time to test/QA' with correlations between 0.074 and 0.273 are considered as the weakest correlations.

\section{Discussion}
This paper explores the adoption and effects of DevOps practices as this has the potential to integrate the software development and software maintenance capabilities of organizations. First, we discuss the adoption of DevOps practices and how this relates to the DevOps maturity. Second, we interpret the impact and effects of DevOps practices. Third, we provide our recommendations for practice and science. Last, we elaborate on the limitations of this study.

\subsection{Adoption of DevOps practices}
Our findings show that 13 out of the 14 identified DevOps practices were adopted by our participants. `Everything as code under version control' was the most adopted practice, which is in line with previous research~\cite{forsgren:2018,ebert:2016,humble:2011}. This practice is key in enabling a fully automated pipeline~\cite{forsgren:2018,lwakatare:2016}, improving the automated and continuous deployments (which show strong correlations with maturity). We believe the high adoption of `everything as code under version control' can be further explained by the trend in containerization of software. In our data we observed that 48\% of our participants use containers to deploy their software. This is in line with the 2018 and 2021 State of DevOps report\cite{google:2018,google:2021}. In the 2018 report organizations who apply containerization are between 1.3 and 1.5 times more likely to be `elite' performers, whereas in the 2021 64\% of the participants indicated that containers are their preferred deployment option.

Whereas we identified `everything as code under version control' as the most adopted practice, this practice is not in the top 4 of practices with the strongest correlations with maturity. Overall, we see that all 14 DevOps practices have a positive correlation against the maturity of DevOps, with most correlations being significant. This indicates that the practices all (more or less) contribute to the maturity of DevOps in organizations.

In Fig.~\ref{fig:practice_usage}, we also observed that `sandboxes for minimum deployment' is the exception and was almost not adopted by participants. This also correlates with the impact of this practice with the maturity of DevOps, where we see the lowest and non-significant correlations for this practice. This could lead to the statement that this practice does not contribute to the maturity of DevOps. However, it could also be the case that sandboxes are being replaced by containers or that the practice itself is not self-explanatory. Additionally, we also observe for `trying to reduce time to test/QA' and `trunk-based development' weak correlations with DevOps maturity. Therefore, we recommend organizations starting with a DevOps transformation focus less on these three practices. From a scientific perspective, we do recommend further investigation of these practices in future research to better understand why these practices correlate less with DevOps maturity.

\subsection{Effects of DevOps practices}
Except for the practices `change-based code reviews' and `configuration management', all DevOps practices correlate positively with organizational performance (supported by Fig.~\ref{fig:cor_prac_perf}). This suggests that these practices can be adopted to improve the performance of organizations. The perceived impact after a DevOps implementation is also positive for all measured dimensions (supported by Fig.~\ref{fig:cor_prac_perf}. However, it is interesting to point out that for two metrics, namely `makes work more fun' and `increases predictability of product delivery' there 25\% of the participants indicate a negative impact. As a DevOps transformation comes with a lot of changes to one`s work, we are not surprised that some participants perceive DevOps as less fun than there work before. However, the indicated negative impact on `increases predictability of product delivery' is somewhat surprising as this refers to some of the main benefits of DevOps~\cite{wiedemann:2019,ghantous:2017}. We argue that this could be do to the fact that there is a learning curve within a DevOps transformation. There could also be the case that these participants were outliers in our dataset. Further research would be necessary to understand how the predictability of product delivery evolves over time.

The least adopted practice `sandboxes for minimum deployment' has the lowest and non-significant correlations with maturity. This practice also does not have strong correlations with organizational performance, except for quality and efficiency. 

As we observe Fig.~\ref{fig:devops_maturity}, the strongest correlations can be found on the quality and efficiency metrics. This is consistent with previous work~\cite{riungu:2016,callanan:2016,lwakatare:2019}. 

DevOps practices have less impact on software delivery performance. However, deployment frequency and deployment time are being affected the most. Deployment frequency, in turn, strongly correlates with most metrics for  organizational performance. Customer satisfaction (corr.: 0.338), for example, is positively impacted, which is in line with the work of Wiedemann et al.~\cite{wiedemann:2019}.

For software delivery performance we observe that the practice of `small and continuous releases' has the biggest effect on both deployment frequency and deployment time. This is not surprising.

Interestingly, we also found negative correlations between DevOps practices and organizational performance or software delivery performance. The practice `change-based code reviews' has five negative correlations. the strongest negative correlation -- while not significant -- suggests that adopting this practice can hinder efficiency. 

The effects of the most widely adopted practice of `everything as code under version control' are significant on two software delivery metrics:  deployment frequency and remediation \%. This corresponds with earlier reported benefits~\cite{forsgren:2018,ebert:2016,humble:2011}.

\subsection{Recommendations for adopting DevOps}
For organizations starting out with DevOps, we recommend adopting `everything as code under version control'. Next to that we suggest to focus on the adoption of `automated testing in environments', `automated continuous deployments', `automated and continuous monitoring', and `configuration management'. These practices have the most positive impact on maturity and performance. Organizations also have to be aware that DevOps is not for everybody, as we found that some participants indicate that work is less fun because of the implementation of DevOps. As we found negative effects for `sandboxes for minimum deployment' and not much adoption among participants, we believe further research is necessary. Thus, we advise to not start with this practice. Additionally, we recommend to use our DevOps practices inventory as a first catalog of DevOps practices, which can be extended and enriched with organization-specific subpractices.

\subsection{Unboxing capabilities}
For science we believe that our method for unboxing a capability provides a pragmatic way to explore the routines and practices of a capability, but also correlating routines and practices to performance and maturity. To unveil the software development and software operations capabilities of organizations, we first did an extensive literature search to understand which practices are present in organization. Then we collected data via a survey among different organizations in different industries and relate this to performance and maturity. This provides us with insights how individual practices contribute to the maturity and how it relates to organizational performance. In literature \cite{salvato2011beyond,molina2020multilevel} different research methods are suggested to get a multi-level perspective on capabilities and their routines and practices. Most of these methods require to collect longitudinal data at one specific organizations in order to break down a capability into all of its routines and practices. This requires the access to organizations over a long period of time, with detailed access. Beyond that, it is difficult to compare results to other organizations to verify and validate how a capability is broken down. Comparison to performance and maturity is more difficult as you can only compare to one organization, where it harder to find patterns.

The downside of our approach is that we can't get more detailed and longitudinal data about the routines, as this would require us to choose for 1 specific case (as Salvato et al. \cite{salvato:2009} did in their research). The level of detail limits us to complete overview of the relationship between capabilities and their routines, but still gives a good indication.  As capabilities tend to be quite stable over time, we do see evolution of capabilities, especially in dynamic settings such as with DevOps or in general with digitization. The lack of longitudinal data, limits our observations about the evolution of capabilities and their underlying practices and routines. This also limits our view on how this evolution impacts the maturity and performance.

\subsection{Limitations}
The first limitation we found is that we primarily investigated the ostensive aspects of the DevOps practices, because we used a survey across different organizations. We did not observe the DevOps practices. This limits the granularity of understanding the details of DevOps practices. Further research could focus on the performative aspects of the practices. One could measure that within a case study.

Another limitation is that we asked our respondents about their perceived organizational performance and effects of DevOps. However, we did not measure the actual organizational performance. For future research it would be interesting to investigate the effects and adoption of DevOps practices on the actual organizational performance, and observing the effects in addition to asking the perceived effects.

While we applied rigor in the setup of this research, designing the survey and collecting/analyzing data, there are limitations to this research. Here we present three types of biases, present particularly in survey-based research: sampling bias, response bias, and non-response bias.

The first bias is related to sampling: the way respondents are chosen to participate. Two roles (Software Developer/Engineer and IT Operations/Infrastructure Engineer) are responsible for more than 50\% of the respondents. This might have lead to self-selection bias: respondents selecting themselves to participate in the research. Self-selection bias was mitigated by sharing the survey in different online communities, consisting of different roles and nationalities.

The second type of bias is the response bias: social desirability in answering questions. We investigated perceived performance, maturity, and impact rather than actual hard data. Participants could provide answers they think are social desirable or be unaware of the performance, maturity, or the impact. The survey was set up anonymous. Therefore, we believe to have mitigated this bias.

The third type of bias is non-response bias: people not participating in the study differ significantly from those who do, resulting in a under-representation. Since the response rate of the survey was 36.5\%, this might have been the case. The input of managers or coaches, for example, takes up less than 15\% of the respondents. Their responses could differ significantly from DevOps practitioners, and can be a good addition for future work.

Another limitation is the lack of similar research on this topic. Although many scholars have given input on the best DevOps practices, their input sometimes originated from single case studies. Although based on relevance and occurrence, the practices chosen for the survey could have a misalignment between theory and practice. Future work could repeat the research with an extended list of practices tools. Future work could also focus on collecting longitudinal data within one organization.

\section{Conclusion}
As DevOps practices are becoming more prominent within the software development and software maintenance capabilities of organizations, we conducted an international survey study to better understand the adoption and effects of DevOps practices on an organization's performance. We introduced an inventory of 14 DevOps practices and ranked their adoption among 123 participants. We found that 13 of the DevOps practices were adopted by most of the participants. Effects have been reported across organizational performance and software delivery performance. Additionally, we also measured the impact of DevOps on the participants themselves.

We conclude that the effects of the DevOps practices are primarily positive on organizational performance. Yet, DevOps is also considered to have negative impact on some participants or organizations, such as participants indicating that work has become less fun or `change-based code reviews' having a negative impact, although not significant, on performance, customer satisfaction, quality, efficiency, and achieving goals. The overview of DevOps practices and their effects can be used for more targeted application of DevOps, obtaining positive effects while minimizing any negative effects. The inventory of practices can be expanded and enriched with future research.

\bibliographystyle{IEEEtran}
\bibliography{bibliography}
\end{document}